\documentclass[12pt]{iopart}
\usepackage{graphicx}

\usepackage{epsfig}
\usepackage{graphicx}
\usepackage{color}
\newcommand{\hA}{\hat{A}}
\newcommand{\ha}{\hat{a}}
\newcommand{\hb}{\hat{b}}
\newcommand{\hain}{\hat{a}_\mathrm{in}}
\newcommand{\haout}{\hat{a}_\mathrm{out}}
\newcommand{\hml}{\hat{m}_\parallel}
\newcommand{\hmp}{\hat{m}_\perp}
\begin{document}

\title{Quantum Theory of All-Optical Switching in Nonlinear Sagnac Interferometers}

\author{Yu-Ping Huang and Prem Kumar}

\address{Center for Photonic Communication and Computing, EECS
Department, Northwestern University, Evanston, IL 60208-3118}


\begin{abstract}
Recently, our group has demonstrated an ultrafast, low-loss, fiber-loop switch based on a nonlinear Sagnac-interferometer design, using which entangled photons were shown to be routed without any measurable degradation in their entanglement fidelity [Hall {\it et al.}, Phys. Rev. Lett. {\bf 106}, 053901 (2011)]. Such a device represents an enabling technology for a rich variety of networked quantum applications. In this paper we develop a comprehensive quantum theory for such switches in general, i.e., those based on nonlinear Sagnac interferometers, where the in-coupling of quantum noise is carefully modeled. Applying to the fiber-loop switch, the theory shows good agreement with the experimental results without using any fitting parameter. This theory can serve as an important guiding tool for configuring switches of this kind for future quantum networking applications.
\end{abstract}

\section{Introduction}
Networked quantum processing hinges on the capability to route photonic signals at high speed, with minimal loss and added in-band noise, and---most importantly---without disturbing their quantum states \cite{NieChu00}. Traditional optical-switching technologies fail to simultaneously satisfy these requirements; see \cite{SwitchNJP11} and references therein for a detailed discussion. Recently, our group has developed an ultrafast, low-loss switch based on a nonlinear Sagnac fiber-loop, using which photonic entanglement was shown to be redirected without any measurable degradation in its fidelity \cite{HalAltKum10,HalAltKumPRL11,PatelNealHall11}. In a nutshell, this switch is an all-fiber implementation of quantum-optical Fredkin gate \cite{FredkinGate,Milburn89,HuangKumar10,HuaKumSwitch11}, which is an important step toward the development of next-generation quantum networking, while harnessing the existing fiber-optic communications infrastructure.

In this paper, we present a quantum theory for the purpose of fully describing the behavior of all-optical switches of this kind, i.e., those based on Kerr-nonlinear Sagnac effect (KNSE). To this end, we develop a comprehensive model starting from a general nonlinear wave equation in (effective) Kerr media, taking into account effects such as group-velocity dispersion, linear propagation loss, cross-phase modulation (XPM), self-phase modulation (SPM), and the in-coupling of quantum noise via background photon emission. To test its validity, we apply the model to describe the fiber-loop switch \cite{HalAltKumPRL11}, where good agreement is found between experimental data and theoretical results that are computed with no fitting parameter. We additionally calculate the switch's quantum-noise level in the presence of spontaneous Raman scattering, which reveals that this switch introduces a negligible amount of in-band quantum noise, in accord with the experimental results. As a whole, the theoretical and experimental results indicate that this switch represents an enabling technology for a rich variety of potent applications such as distributed quantum computing \cite{Qist_Quan_Comp_97,QIP_Rev_02}, quantum communication using hyper-entanglement \cite{HyperEntanglement05}, and quantum receiver \cite{Guha11}. The theory developed here can prove important for configuring the KNSE-based switches for such applications.

The paper is organized as follows. In section \ref{Sec_SD} we describe the KNSE-switch setup in general as well as that based on a fiber-loop design. In section \ref{Sec_Model}, we develop a general theory for the quantum description of such switches. Then in section \ref{Sec_FL}, we apply the theory to the fiber-loop switch and compare with the experimental results. Lastly, in section \ref{Sec_CL} we conclude with a brief discussion.

\section{Switch Setup}
\label{Sec_SD}

\begin{figure}
   \centering \epsfig{figure=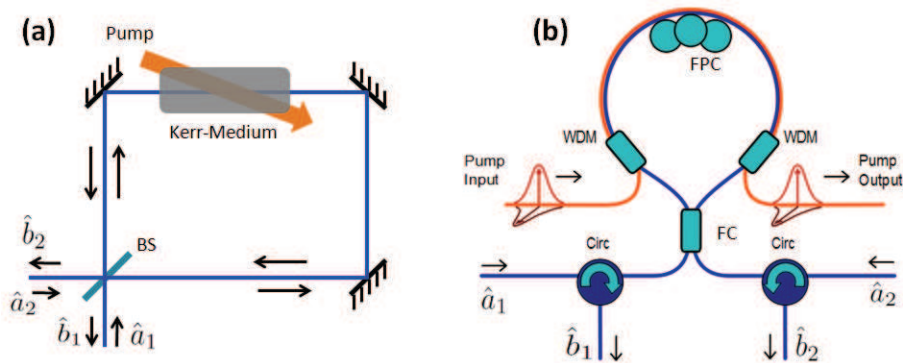,width=12cm}
   \caption{(a) A schematic showing the general setup of the KNSE switches. (b) A fiber-loop implementation of such switches. BS: 50/50 beamplitter, WDM: wavelength-division multiplexer, FPC: fiber polarization controller, Circ: circulator, FC: 50/50 fiber coupler. Note that the design of the present fiber-loop switch is modified from our previous experimental setup in that there are now two inputs, instead of one, for the signal waves \cite{SwitchNJP11,HalAltKumPRL11}.}  \label{fig1}
\end{figure}

The general setup of the KNSE switches is schematically shown in Fig.~\ref{fig1}(a). It is essentially a Sagnac-loop interferometer incorporating an intra-loop Kerr nonlinear medium. The Kerr medium can be made directly of $\chi^{(3)}$ materials, such as silicon microrings \cite{Switch-Mircroring-2004} and silica fibers \cite{Mor88,FibLooRel89}, or composed effectively of cascaded $\chi^{(2)}$ materials, such as lithium-niobate waveguides \cite{IroAitArn93,KanItoAso99}. There are two inputs and two outputs for the signal waves (``target'' beams), whose time-domain annihilation operators are labeled $\ha_{1,2}(t)$ and $\hb_{1,2}(t)$, respectively, satisfying $[\ha_{j}(t),\ha^\dag_{k}(t')]=[\hb_{j}(t),\hb^\dag_{k}(t')]=\delta_{jk}\delta(t-t')$. The input signal waves split first on a 50/50 beamsplitter into two components, which then counter-propagate along clockwise and counterclockwise directions in the Sagnac loop. When no pump (``control'' beam) is applied, the two components acquire the same amount of phase shift during their propagation in opposite directions, so that when recombining at the beamsplitter, they exit from the entering port, i.e., they are reflected by the loop. In this case, the switch is effectively in a ``bar'' state. When pump pulses are applied along one propagation direction, however, the clockwise and anticlockwise propagating signal waves experience different phase shifts via XPM as they pass through the Kerr medium. When the difference between the two phases is an odd-integer multiple of $\pi$, the signal waves will exit totally from the other port, i.e., they are transmitted by the loop. The switch is now in an effective ``cross'' state. Note that because XPM is usually polarization dependent, achieving polarization-insensitive switching of the signal waves requires unpolarized pump pulses. This can be accomplished by temporally overlapping two orthogonally polarized pumps of identical power but with slightly different wavelengthes \cite{BulVei93}.

A fiber-loop implementation of such a KNSE switch is drawn in Fig.~\ref{fig1}(b), in which a spool of optical fiber simultaneously constitutes the Sagnac-loop path and acts as the Kerr medium \cite{FibLooRel89,Blow90}. In this design, the input and output signal waves are separated by use of optical circulators, while the pump pulses are coupled into and out of the fiber loop via two wavelength-division multiplexers (WDMs). The switch is initially configured to be in the ``bar'' state by adjusting an intra-loop fiber polarization controller. By multiplexing strong pump pulses into the fiber loop, a phase difference is created between the counter-propagating components of the signal waves through XPM via the fiber's Kerr nonlinearity. Tuning the pump power such that the phase difference reaches $\pi$, the switch goes into in a ``cross'' state, thereby accomplishing a switching operation.

For the fiber-loop switch shown in Fig.~\ref{fig1}(b), there are two conditions that are critical for simultaneously achieving high switching contrast and low added in-band noise. First, the pump pulses and the signal waves must have a group velocity difference large enough for a pump pulse to walk completely through the signal's temporal extent, so that the latter can gain a uniform phase shift \cite{Blow90,AOS_rev_97}. Second, the pump wavelength must be chosen such that it does not produce a significant amount of background photons in the signal band \cite{KarDouHau94,cpx,Agr07}. In our experiment, we used a 1550-nm wavelength for the pump pulses and 1310 nm for the signal waves, the combination of which led to simultaneous satisfaction of these two requirements \cite{SwitchNJP11}.

\section{Quantum Model}
\label{Sec_Model}
We now develop a quantum theory for the KNSE switches whose general setup is sketched in Fig.~\ref{fig1}(a). Due to the similarity of Sagnac interferometers with Mach-Zehnder (MZ) interferometers, the present theory can be easily modified to describe the latter kind of switches as well \cite{SanMil92}. Our first step is to analyze the propagation of the pump pulses and the signal waves inside the Kerr medium. We start from the most general Heisenberg equation of motion for the slowly-varying operators $\{\hA_j(z,t)\}$ of the optical fields propagating along the $z$ direction in a guided $\chi^{(3)}$ medium \cite{Drum96}:
\begin{eqnarray}
\label{eq14}
\frac{\partial\hA_j(z,t)}{\partial z}&=&i\sum_{k} \left[\int^t_{-\infty} R^{(1)}_{jk}(t-t')\hA_k(z,t')dt' +\sqrt{\hbar\omega_0}~\hat{m}_{jk}(z,t)\hA_k(z,t)\right] \nonumber\\
 &+&i\sum_{klm}\int^t_{-\infty} R^{(3)}_{jklm}(t-t') \hA^\dag_k(z,t') \hA_l(z,t')\hA_m(z,t)dt'.
\end{eqnarray}
Here, the subscripts $j,k,l,m$ denote the polarization directions of the optical waves; $R^{(1)}_{jk}$ and $R^{(3)}_{jklm}$ represent the linear and third-order nonlinear responses of the medium, respectively, to the applied optical fields; $\omega_0$ is the optical carrier frequency; and $\{\hat{m}_{jk}\}$ are operators describing the in-coupling of quantum noise, which, in the case of the fiber-loop switch, is dominated by the physical process of spontaneous Raman scattering.

For the KNSE switches, the pump pulses are usually strong in order to produce the desired $\pi$ phase difference with weak (effective) Kerr nonlinearities. They can thus be treated as classical mean fields, allowing us to expand the optical-field operators as
\begin{equation}
\label{eq13}
    \hA(z,t) \vec{\epsilon}=A_1(z,t) \vec{\epsilon}_1~e^{-i\omega_1 t}+A_2(z,t) \vec{\epsilon}_2~ e^{-i\omega_2 t}+\ha_s(z,t) \vec{\epsilon}_s~ e^{-i\omega_s t},
\end{equation}
where $A_1$ and $A_2$ are the amplitudes of the orthogonally polarized pump pulses whose optical frequencies relative to $\omega_0$ are  $\omega_1$ and $\omega_2$, respectively; $\ha_s$ is the annihilation operator for the signal wave whose relative frequency is $\omega_s$; and the vectors $\vec{\epsilon}$ and $\vec{\epsilon}_{1,2,s}$ denote the polarization directions of the respective optical field. Inserting Eq.~(\ref{eq13}) into Eq.~(\ref{eq14}) and expanding the linear response of the Kerr medium up to the second order, we arrive at the following classical wave equation that dominates the pump-pulse dynamics:
\begin{eqnarray}
\label{eq8}
\frac{\partial A_j(z,t)}{\partial z}&=&-\alpha A_j(z,t)-\beta^{(1)}_j \frac{\partial A_j(z,t)}{\partial t}-\frac{i\beta^{(2)}_j}{2} \frac{\partial^2 A_j(z,t)}{\partial t^2} \nonumber \\
&+& i\left(\varrho P_\parallel(z,t)+\varsigma P_\perp(z,t)\right) A_j(z,t),~~~~j=1,2.
\end{eqnarray}
Here, $\alpha$ is a linear-loss coefficient. $\beta^{(1)}_j=\int^\infty_0 d\tau R^{(1)}_{jj}(\tau)\tau$ and $\beta^{(2)}_j=\int^\infty_0 d\tau  R^{(1)}_{jj}(\tau)\tau^2/2 $ are related to the group velocity and the group-velocity dispersion of the pump pulses, respectively. $\varrho=\int^\infty_0 d\tau R^{(3)}_{jjjj}(\tau)/\hbar\omega_0$ is the SPM coefficient and $\varsigma=2\int^\infty_0 d\tau R^{(3)}_{jkjk}(\tau)/\hbar\omega_0$ ($j\neq k$) is the coefficient for XPM. $P_\parallel$ and $P_\perp$ are the instantaneous powers of the pump pulses in the parallel and perpendicular polarization directions with respect to $\vec{\epsilon}_j$. In arriving at Eq.~(\ref{eq8}), we have assumed that the response time of Kerr nonlinearity is much shorter than the pump-pulse duration, which is indeed the case for typical Kerr media. For example, for telecommunication fibers, the response time is about $0.05$ ps \cite{LinAgr06}, which is much shorter than the typical pump-pulse duration used ($>1$ ps). Lastly, we have also neglected a term of the form $A^\ast_j A_k A_k$ ($j\neq k$), describing coherent coupling between the orthogonally-polarized pump pulses, because it corresponds to an energy non-conserving process.

Equation (\ref{eq8}) can be solved analytically under certain circumstances \cite{Agr07} or numerically in general to obtain the pump-pulse amplitude at any $(z,t)$ point inside the Kerr medium. Using the solution obtained in either way, the Heisenberg equation of motion for the signal-field operator $\ha_s$ can be linearized to give
\begin{eqnarray}
\label{eqAs}
     & & \left[\frac{\partial}{\partial z}+\frac{1}{v_s} \frac{\partial}{\partial t}\right]\ha_s(z,t)= i [\xi_{\parallel} P_{\parallel}(z,t)
    +\xi_\perp P_\perp(z,t)] \ha_s(z,t)\nonumber \\
    & &~~~~+i\sqrt{\hbar\,\omega_s}[\hat{m}_{\parallel}(z,t) A_{\parallel}(z,t)+\hat{m}_{\perp}(z,t) A_{\perp}(z,t)]+\hat{\varepsilon}.
\end{eqnarray}
Here, $v_s$ is the group velocity of the signal wave and $\omega_s$ is its center frequency. $\xi_{\parallel}$ and $\xi_{\perp}$ measure the strength of the XPM produced by the pump pulses in parallel and perpendicular polarizations, respectively, relative to the signal wave. $\hml$ and $\hmp$ are the operators describing the in-coupling of quantum noise produced by the pump pulses in parallel ($A_\parallel$) and perpendicular ($A_\perp$) polarization directions, respectively. $\hat{\varepsilon}$ is a vacuum-noise operator whose commutation relation is determined such that the commutation relation for the signal wave is conserved at any position and time inside the Kerr medium \cite{BioKarHau94,Raman02}. In arriving at Eq.~(\ref{eqAs}), we have assumed that the Kerr medium is shorter than any of the characteristic lengths associated with propagation loss and group-velocity dispersion of the signal wave. In addition, we have again used the fact that the time response of the Kerr nonlinearity is much faster than any other time scale of the system. Finally, we have neglected background photon-pair emission via spontaneous four-wave mixing (SFWM) of the pump waves, as this process is highly phase-mismatched in our systems. Take the fiber-loop switch as an example, the phase-mismatch for SFWM involving the pump pulses in the 1550-nm C-band and the signal waves in the 1310-nm O-band is about $1.5\times 10^3$ m$^{-1}$. Consequently, the probability to create photon pairs via such SFWM is suppressed by a factor of $\sim 10^{10}$ in a 100-m long fiber, as compared to the phase-matched case.

Equation (\ref{eqAs}) can be solved conveniently in a moving frame that travels with the signal wave. For a Kerr medium of length $L$, we find after some algebra that the output annihilation operator $\haout$ of the signal wave at $z=L$ can be written as
\begin{eqnarray}
\label{haout}
    \hat{a}_{\mathrm{out},\pm}(t)=e^{i\Phi_\pm(t)} \hain(t-L/v_s)+\hat{\eta}_\pm(t)+\hat{\varepsilon}_\pm(t).
\end{eqnarray}
Here, $\hain$ is the input operator of the signal wave at $z=0$. The subscript $+$ ($-$) indicates that the signal copropagates (counter-propagates) with the pump pulses. $\Phi_\pm$ are the XPM phase shifts given by
\begin{eqnarray}
\label{EqPs}
    \Phi_\pm (t)&=&\int^L_0 \left[\xi_\parallel P_\parallel(z,t-L/v_s+z \beta_\pm)+\xi_\perp P_\perp(z,t-L/v_s+z \beta_\pm)\right]dz,
\end{eqnarray}
where $\beta_{\pm}=1/v_s\mp 1/v_p$ ($v_p$ is the group velocity of the pump pulses) is a ``walk-off'' parameter measuring the group-velocity difference between the signal wave and the pump pulses. The operators
\begin{equation}
\label{Eqeta}
      \hat{\eta}_\pm(t)= i\sqrt{\hbar\omega_s} \int^L_0\!\!\! \sum_{j=\parallel,\perp} \!\!\! \hat{m}_j(z,t+(z-L)/v_s) A_j(z,t-L/v_s+z \beta_\pm)dz
\end{equation}
account for the in-coupling of quantum noise via background photon emission. $\hat{\varepsilon}_\pm$ are the vacuum-noise operators whose commutation properties are chosen so as to preserve the commutation relations of the output signal-field operators.

Using Eq.~(\ref{haout}) and following standard procedures \cite{ManWol95}, we arrive at the following relation for the switch input ($\ha_{1,2}$) and output ($\hat{b}_{1,2}$) operators:
\begin{equation}
\label{eq2}
    \left(
      \begin{array}{c}
        \hat{b}_1 \\
        \hat{b}_2 \\
      \end{array}
    \right)= 
    e^{i\varphi(t)}e^{-\ell_s} \left(\!\!
      \begin{array}{cc}
        \cos\theta(t) & i\sin\theta(t) \\
         i\sin\theta(t)& \cos\theta(t)  \\
      \end{array}
    \right) 
        \left(
      \begin{array}{c}
        \hat{a}_1 \\
        \hat{a}_2 \\
      \end{array}
   \!\! \right)
+e^{-\ell_r}\left(\!\!
      \begin{array}{c}
        \hat{\eta}_1\\
        \hat{\eta}_2\\
      \end{array}
    \right)
    +   \left(\!\!
      \begin{array}{c}
        \hat{\varepsilon}_1 \\
        \hat{\varepsilon}_2 \\
      \end{array}
    \right),
\end{equation}
where
\begin{eqnarray}
    \varphi(t)&=&[\Phi_+(t) +\Phi_-(t)]/2, \\
\label{eq12}
     \theta(t)&=&[\Phi_+(t)-\Phi_-(t)]/2, \\
     \hat\eta_{1,2}&=&(\hat{\eta}_+\pm \hat{\eta}_-)/\sqrt{2}.
\end{eqnarray}
In Eq.~(\ref{eq2}), $\ell_s$ and $\ell_r$ are two phenomenological parameters introduced to account for the total propagation losses of the signal and the background photons, respectively, as they travel through the switch. Such losses can occur in the Kerr medium, on the beamsplitter, along the Sagnac-loop path, etc. $\hat{\varepsilon}_{1,2}$ are, again, the vacuum-noise operators whose properties are determined such that the commutation relations of the switch-output operators $\hb_{1,2}$ are preserved.

Physically, the terms $\hat{\varepsilon}_{1,2}$ and $\hat{\eta}_{1,2}$ in Eq.~(\ref{eq2}), together, describe in-coupling of phase noises to the signal band via the spontaneous processes. Such noises will have substantial effects when the width of the pump and the signal pulses are comparable with the response time of the nonlinear medium, i.e., in the so-called slow-response regime \cite{Shapiro06,Shapiro07}. In the opposite regime, on the other hand, where the response time is much shorter, these noises can be made negligibly small, for example, by choosing appropriate wavelengths for the pump and the signal waves, as was done in our experiments \cite{SwitchNJP11,HalAltKumPRL11}. This also means that there will be an ultimate limit on the shortest possible switching window that can be achieved with such nonlinear devices; for the fiber-loop switches it will be on the order of 100 fs.

Equation (\ref{eq2}) represents the main result of the present theory, which clearly shows that the KNSE switches are coherent, four-mode devices when the signal-photon loss and the added background noise are negligible. Using such devices, optical signals can be totally switched without their quantum states being disturbed for $\theta=\pi/2$. They can also be coherently split or multiplexed for $\theta=\pi/4$.

The input-output transformation matrix in Eq.~(\ref{eq2}) is quite general; it is applicable to any choice of (effective) Kerr medium and pump pulses. To gain some analytic insights, in the following we consider a special case where the group-velocity dispersion is negligible for the pump pulses, i.e., $\beta_j^{(2)}\simeq 0$. Consider Gaussian pump pulses whose input amplitude profile reads
\begin{equation}
    A_{0}(t)=\sqrt{\frac{P_0}{\hbar \omega_0}} e^{-t^2/2\sigma^2}.
\end{equation}
Using Eq.~(\ref{eq8}), we obtain
\begin{equation}
\label{eqAp}
    A_{j}(z,t)\approx A_{0}(t-z/v_p)e^{i\phi(t-z/v_p)z-\alpha z},
\end{equation}
where $\phi$ is the total phase shift resulting from both the SPM and the XPM processes, and is given by
\begin{equation}
\label{eq9}
    \phi(t-z/v_p)=\varrho P_{\parallel}(t-z/v_p)e^{-2\alpha z}+\varsigma P_\perp(t-z/v_p)e^{-2\alpha z}.
\end{equation}
With the result in Eq.~(\ref{eqAp}), we then analytically integrate Eq.~(\ref{EqPs}) to obtain the cross-phase shifts as
\begin{eqnarray}
\label{eq7}
   \Phi^G_\pm(t) &=& P_0 (\xi_\parallel+\xi_\perp)\mu^G_\pm,
\end{eqnarray}
where
\begin{equation}
\label{eq3}
    \mu^G_+=\frac{1-\exp(-2L\alpha)}{2\alpha}\exp\left[-\frac{(v_s t-L)^2}{2\sigma^2 v_s^2}\right]
\end{equation}
for $v_s=v_p$, and
\begin{eqnarray}
\label{eq4}
    \mu^G_\pm&=&\frac{\sqrt{\pi}\sigma }{2\beta_{\pm}} \exp\left[\frac{\alpha (\alpha \sigma^2+2 t -2L\beta_{\pm}/v_s)}{\beta_\pm}\right]\nonumber\\
    & & \times \left[\mathrm{Erf}\left(\!\frac{L-v_s t+\alpha \sigma^2 v_s/\beta_\pm}{v_s \sigma} \!\right)+\mathrm{Erf}\left(\!\frac{v_p t\mp L+\alpha \sigma^2 v_p/\beta_\pm}{v_p \sigma} \!\right)\!\right]
\end{eqnarray}
for all other cases.

From Eq.~(\ref{eq4}), $\Phi_+$ is proportional to $|v_s-v_p|^{-1}$ while $\Phi_-$ is proportional to $(v_s+v_p)^{-1}$. Thus, in typical Kerr media with $v_s\approx v_p$, we have $\Phi_+(t)\gg \Phi_-(t)$, i.e., the cross-phase shift gained by the component of the signal wave copropagating with the pump pulses is much larger than that experienced by the counter-propagating component. This feature of directional cross-phase shifts produces the desired phase difference between the clockwise- and anticlockwise-propagation directions in the Sagnac loop, which eventually leads to switching of the signal wave.



Equations (\ref{eq12}) and (\ref{eq7}) show that when $v_s=v_p$, i.e., when the signal and the pump pulses propagate in lockstep
through the entire nonlinear medium, the phase difference gained by the signal wave via XPM is nonuniform across its temporal extent. As a result, it can only be partially switched \cite{Eiselt92}. In order to achieve high-contrast switching then requires either the use of temporally square-shaped pump pulses or the condition of $v_p\neq v_s$. As square pulses are inconvenient to operate with in practice because of instantaneous rise and fall times, we consider using the latter condition for which the signal wave slides relative to the pump pulse as they copropagate in the Kerr medium. When this occurs, as seen from Eq.~(\ref{eq4}), a time window with a flat cross-phase shift (i.e., constant over time) can be created as long as $L>2\sigma/|\beta_+|$. Physically, in this case the signal wave within that window will have time to completely `walk through' the pump pulse, thereby creating a uniform cross-phase shift that depends only on the pump-pulse energy, not its shape. A detailed intuitive explanation can be found in Fig.~5 of Ref. \cite{SwitchNJP11}.

In the special case when the pump-propagation loss is negligible, i.e., $\alpha\approx 0$, a total-transmission window will be created for the switch when the pump-pulse energy $E\equiv 2\sqrt{\pi}\sigma P_0$ (summing over the two orthogonal polarizations) satisfies
\begin{equation}
\label{eq6}
    E=\frac{2\pi|\beta_+|}{\xi_\parallel+\xi_\perp}.
\end{equation}
We note that while this result is derived here assuming a Gaussian profile for the pump pulses, it can actually be applied to any pulse shape. In other words, a total switching window can be realized as long as the energy contained in a pump pulse meets this condition, regardless of the pulse-shape details. This feature makes the KNSE switches quite robust against optical imperfections in handling of the pump pulses. The total-transmission window is centered at
\begin{equation}
   t_c=L\beta_-/2,
\end{equation}
with a full-width at half-maximum of
\begin{equation}
\label{tw}
    \tau_w=L|\beta_+|-2\sigma \mathrm{Erf}^{-1}\left(\pi/4\right).
\end{equation}
This result shows that the width of the switching window is determined by the physical length and the linear-optical property of the Kerr medium. Its rise and fall times, on the other hand, are determined by the width of the pump pulses used. A narrower switching window can be obtained by using a smaller group-velocity difference between the signal and the pump waves and/or longer-duration pump pulses.

\section{Fiber-loop Switch}
\label{Sec_FL}
Using the general theory developed in the previous section, we next study specifically the fiber-loop switch demonstrated recently in our lab \cite{HalAltKum10,HalAltKumPRL11}. For such a system, the in-coupling of quantum noise is predominantly through the physical process of spontaneous Raman emission, with the corresponding noise operators obeying \cite{LinAgr06,BioKarHau94}
\begin{eqnarray}
\label{eq10}
    [\hat{m}_{jk}(z,\tau),\hat{m}_{lf}(z',\tau')]=i\delta(z-z')[R^{(3)}_{lfjk}(\tau'-\tau)-R^{(3)}_{jklf}(\tau-\tau')],
\end{eqnarray}
where $R^{(3)}_{jklf}$ is the fiber nonlinear-response function given by \cite{Agr07}
\begin{eqnarray}
    R^{(3)}_{jklf}(\tau-\tau')&=& \gamma(1-f_R)\delta(t)(\delta_{jk}\delta_{lf}+\delta_{jl}\delta_{kf}+\delta_{jf}\delta_{kl})/3
\nonumber\\
& & +\gamma f_R R_a(\tau)\delta_{jk}\delta_{lf}+\gamma f_R R_b(\tau)(\delta_{jl}\delta_{kf}+\delta_{jf}\delta_{kl})/2.
\end{eqnarray}
Here, $R_{a,b}(\tau)$ are the isotropic and anisotropic Raman response functions, respectively; $\gamma$ is the nonlinear coefficient of the fiber as defined in \cite{Agr07}; $f_R$ represents the fractional contribution of the Raman response to the fiber nonlinearity, which equals 0.18 for typical single-mode fibers; and the XPM coefficients are given by
\begin{eqnarray}
\label{eq11}
    \xi_{\parallel}&=& 2\gamma+f_R\gamma[\tilde{R}_a(\Omega)+\tilde{R}_b(\Omega)-1], \\
    \xi_{\perp}&=& 2\gamma/3+f_R\gamma[\tilde{R}_a(\Omega)+\tilde{R}_b(\Omega)/2-2/3],
\end{eqnarray}
with $\tilde{R}_{a,b}(\Omega)=\int^\infty_{-\infty} R_{a,b}(\tau) e^{i\Omega\tau} d\tau$ and $\Omega$ being the detuning of the signal wave from the pump wave.

\begin{figure}
   \centering \epsfig{figure=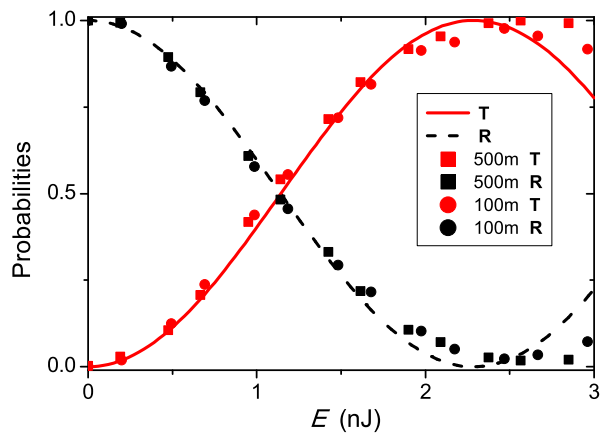,width=8cm}
   \caption{The switch transmission ($T$) and reflection ($R$) probabilities as functions of the pump-pulse energy $E$. The symbols show the experimental results \cite{SwitchNJP11, HalAltKumPRL11} for the fiber-loop switch consisting of 100-m and 500-m long fibers, respectively, while the solid and dashed curves are calculated via Eqs.~(\ref{eq7}) and (\ref{eq4}) without using any fitting parameter. Note that the theory curves overlay on top of each other for the two fiber-loop lengths.}
   \label{fig2}
\end{figure}

As a test of our theory, in the following we simulate the behavior of the fiber-loop switch using standard fiber parameters \cite{Agr07}. In particular, we set $\beta_+$ to be 2.1 ps/m as obtained by integrating the dispersion curve of the standard single-mode fiber (SMF-28) used \cite{Agr07,MHD10}. The results for the signal transmission and reflection as functions of the pump-pulse energy $E$ are plotted in Fig.~\ref{fig2}. As shown, the theory curves are in good agreement with the experimental data without the use of any fitting parameter. The deviation of the theory curves from the measured results at high pump energies, however, is likely due to the nonlinear loss suffered by the pump pulses in that regime. Such nonlinear loss can arise from stimulated Raman scattering, an effect which has not been taken into account in the present model.

\begin{figure}
   \centering \epsfig{figure=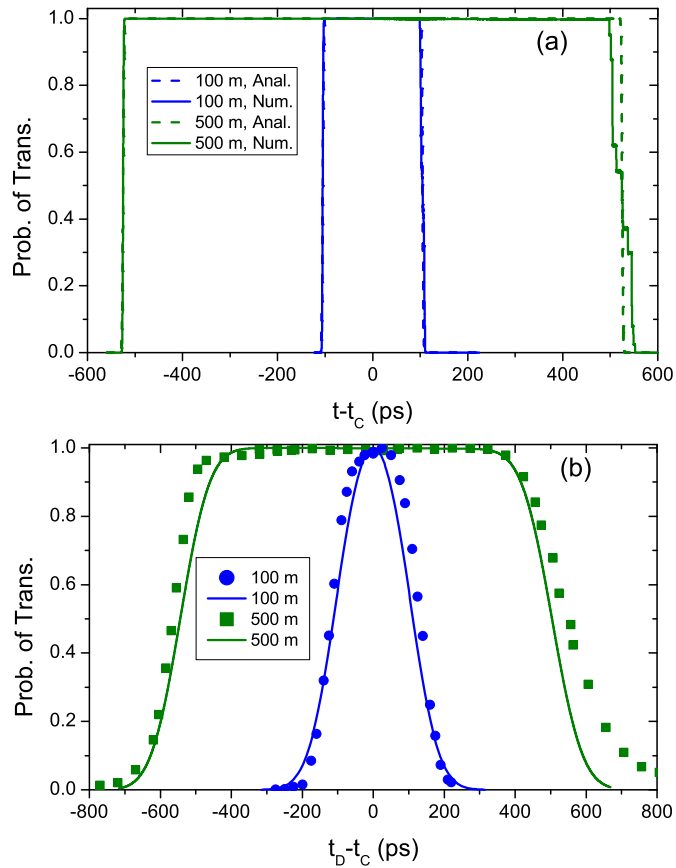,width=9cm}
   \caption{(a) The switching windows for the fiber-loop switch consisting of 100-m and 500-m long fibers, respectively, showing the analytical results (``Anal.'') using Eqs.~(\ref{eq2})--(\ref{eq4}) and numerical results (``Num.'') obtained by directly solving Eqs.~(\ref{eq8}) and (\ref{eqAs}). (b) The probability for 200-ps signal photons to be switched as a function of their delay time $t_D$ with respect to the pump pulses, where the symbols are experimental data \cite{SwitchNJP11, HalAltKumPRL11} and the curves are theoretical results.}
   \label{fig4}
\end{figure}

In Fig.~\ref{fig4}(a), we plot the simulation results for the switching windows. As shown, the analytical results obtained from Eqs.~(\ref{eq2})--(\ref{eq4}) turn out to agree well with the exact results obtained by numerically solving Eqs.~(\ref{eq8}) and (\ref{eqAs}) via the split-step Fourier method \cite{SSFM1,SSFM2}. There is, however, a slight disagreement between the two at the falling (right) edge of the switching window for the 500-m switch. We attribute this discrepancy to the fact that the analytical formulas in Eqs.~(\ref{eq2})--(\ref{eq4}) are derived by neglecting the dispersion of the pump pulses, an assumption that is not necessarily valid in the present system. This is because while the dispersion length of the fiber is quite long (about 1.3 km for the 5-ps pump pulses used), the nonlinear length (about 1 m) is much shorter than the fiber-loop length \cite{Agr07}. As a result, the pump pulses are significantly broadened and distorted as they propagate through the fiber due to the effects of SPM and XPM. Such broadened and distorted pump pulses affect only the falling behavior of the switching window, as depicted in Fig.~\ref{fig4}(a). The rising (left) edge of the switching window, on the other hand, remains unaffected, as the signal wave in that part of the switching window manages to walk off the pump pulse before any significant pump-pulse broadening takes place.

\begin{figure}
   \centering \epsfig{figure=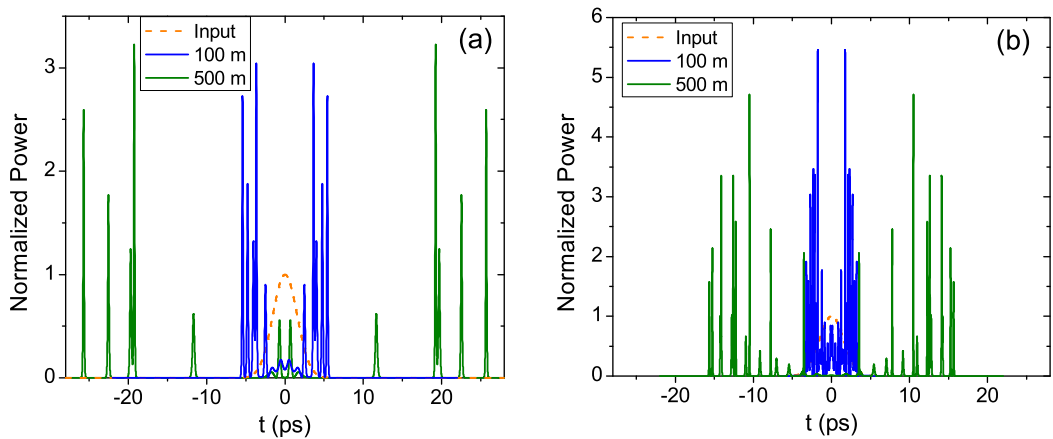,width=14cm}
   \caption{The input and evolved pump-pulse shapes after traveling over distances of 100 and 500 m, respectively, in standard single-mode fiber (a) and non-zero dispersion-shifted fiber (b). In both figures, the fiber mode-field diameters are chosen to be 10 $\mu$m. Note that the sharp peaks shown above are likely not physical as the simulation has not taken into account higher-order effects such as stimulated Raman scattering and third-order dispersion, which are expected to play a non-negligible role in the parameter regime considered here.}
   \label{fig5}
\end{figure}

To examine the extent of pump-pulse broadening, we simulate the evolution of a pair of orthogonally-polarized pump pulses propagating in the fiber using Eq.~(\ref{eq8}). For the input pump pulses, we consider a Gaussian profile of 5-ps FWHM and a total energy of 2.5 nJ. Figure \ref{fig5}(a) shows the results for standard single-mode fiber (SMF-28), wherein the pump pulses spread to over 10 (50) ps after traveling through 100 (500)~m of fiber. 
Such pulse-broadening effect can be reduced by replacing SMF-28 in the Sagnac loop with a non-zero dispersion-shifted fiber (NZDSF), which has a much smaller group-velocity dispersion coefficient for the pump pulses in the telecom band. In Fig.~\ref{fig5}(b) we plot the resulting shapes of the pump pulses traveling in a piece of NZDSF, where all the parameters are the same as in Fig.~\ref{fig5}(a) except that $\beta^{(2)}$ is $-3$ ps$^2$/km for the NZDSF instead of $-20$ ps$^2$/km for SMF-28. As shown, the pulse spreading is reduced by about half, i.e., to around 6 ps and 26 ps for the 100-m and 500-m cases, respectively.

In Fig.~\ref{fig4}(b), we plot the probability for a 200-ps Gaussian signal pulse to be switched as a function of its delay time $t_D$ relative to the pump pulse. The simulated curves are computed by convoluting the signal-pulse profile with the numerical results for the switching window shown in Fig.~\ref{fig4}(a). Again, the theory curves turn out to agree well with the experimental data without any fitting parameter. The fact that the measured switching window is slightly wider than the theoretical prediction is due to the fact that the actual signal pulses used for this measurement had long, non-Gaussian tails \cite{SwitchNJP11}.

\begin{figure}
   \centering \epsfig{figure=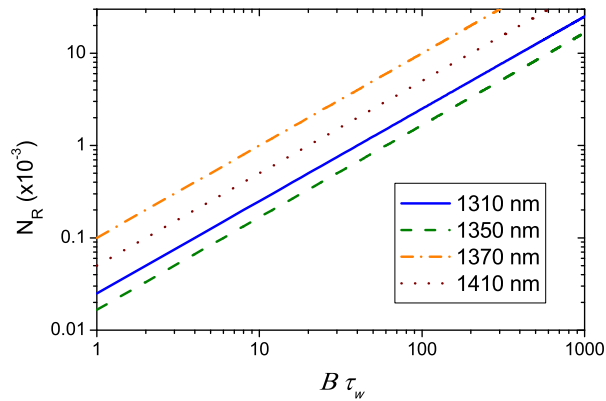,width=8cm}
   \caption{$N_R$ as a function of $B\tau_w$, computed for pump pulses centered at 1550 nm and for various signal wavelengths.}
   \label{fig3}
\end{figure}

Lastly, we study the level of quantum noise introduced by the fiber-loop switch as a result of spontaneous Raman scattering from the pump pulses. Using Eqs.~(\ref{Eqeta}) and (\ref{eq10}), we obtain the flux of the Raman photons generated in the signal band $B$ (in Hertz) to be
\begin{eqnarray}
\label{eq5}
    I_R(t)=\gamma f_R P_0 B \mu_+(t)n_\mathrm{th}(\Omega)[g_a(\Omega)+3g_b(\Omega)/2],
\end{eqnarray}
where $n_\mathrm{th}(\Omega)=[\exp(\hbar\Omega/k_B T)-1]^{-1}$ is the mean number of phonons in a reservoir at equilibrium temperature $T$ with $k_B$ being the Boltzman constant, and $g_a$ and $g_b$ are the Raman-gain coefficients that are related to the isotropic and anisotropic Raman response functions, respectively, by $g_{a,b}=2 \Im(\tilde{R}_{a,b})$. Inserting Eq.~(\ref{eq6}) into Eq.~(\ref{eq5}), the average total number of Raman photons generated in the signal band during the total-switching window is given by
\begin{equation}
    N_R=\gamma f_R B \tau_w n_\mathrm{th}\frac{\pi}{4}\frac{2g_a+3g_b}{\xi_\parallel+\xi_\perp}.
\end{equation}
This result shows that the level of Raman-effect generated noise, which is an inherent property of the fiber-loop switch, depends on the optical properties and the temperature of the fiber. As an example, in Fig.~\ref{fig3} we plot $N_R$ as a function of $B \tau_w$ for a room-temperature fiber-loop switch with a pump wavelength of 1550 nm and for various signal wavelengths. As shown, for a switching window of $\tau_w\sim 1/B$, the total number of Raman photons $N_R$ is about $2\times 10^{-5}$ for signal waves at 1310 nm, which is in accord with our experimental results in \cite{SwitchNJP11}. Even for an extremely long switching window $\tau_w= 1000/B$, the total number of Raman photons $N_R$ is less than $0.04$ for the plotted signal wavelengths. These results confirm that the fiber-loop switches can be operated at very low quantum-noise levels, and are hence suitable for quantum applications.

Consider, for example, the switching of the signal photon of an entangled photon-pair via the experimental setup of Fig.~7 in Ref. \cite{SwitchNJP11}. In this application, the quantum noise introduced by the switch will lead to a degradation of the signal photon's entanglement fidelity with its partner (idler) photon. The amount of degradation can be easily computed using Eq. (8) for any input state of the photon-pair. Particularly, for a perfectly entangled input state, we find that the entanglement fidelity at the output of the switch will be reduced to $1-N_R/2$ when $N_R$ is small (the $1/2$ factor accounts for the fact that only half of the Raman-scattered photons exit from the same output port as the signal photons). For the specific parameters in \cite{SwitchNJP11}, this gives an output entanglement fidelity of 99.999\%, which is consistent with the experimental findings in \cite{SwitchNJP11} where no measurable degradation of entanglement was observed.

\section{Conclusion and Discussion}
\label{Sec_CL}
In conclusion, we have presented a quantum theory for all-optical switching based on the Kerr-nonlinear Sagnac effect. Starting from a general nonlinear wave equation applicable in (effective) Kerr media, we have developed a comprehensive model for such switching devices taking into account the in-coupling of quantum noise. Our analysis shows that such switches are coherent single-mode devices that can be utilized for routing or splitting of quantum signals. Applying to the fiber-loop switch recently demonstrated in our lab \cite{SwitchNJP11, HalAltKumPRL11}, the model turns out to agree well with the experimental data without the use of any fitting parameter. Therefore, the quantum theory presented here can serve as an important guiding tool for configuring switches of this kind in a rich variety of networked quantum applications, such as distributed quantum computing, quantum communications, and quantum receivers.

\section{Acknowledgements}
We would like to thank J. B. Altepeter and N. N. Oza for helping with this manuscript. This research was supported in part by the Defense
Advanced Research Projects Agency under
Grant No. W31P4Q-09-1-0014 and by the U.S. Air
Force Office of Scientific Research under Grant
No. FA9550-09-1-0593.

\section{References}
\bibliographystyle{unsrt}

\end{document}